# D-SPACE4Cloud: A Design Tool for Big Data Applications

Michele Ciavotta*, Eugenio Gianniti**, and Danilo Ardagna* * *

Dipartimento di Elettronica, Informazione e Bioingegneria,
Politecnico di Milano

**Abstract** The last years have seen a steep rise in data generation worldwide, with the development and widespread adoption of several software projects targeting the Big Data paradigm. Many companies currently engage in Big Data analytics as part of their core business activities, nonetheless there are no tools and techniques to support the design of the underlying hardware configuration backing such systems. In particular, the focus in this report is set on Cloud deployed clusters, which represent a cost-effective alternative to on premises installations. We propose a novel tool implementing a battery of optimization and prediction techniques integrated so as to efficiently assess several alternative resource configurations, in order to determine the minimum cost cluster deployment satisfying Quality of Service constraints. Further, the experimental campaign conducted on real systems shows the validity and relevance of the proposed method.

## 1 Introduction

Nowadays, the Big Data adoption has moved from experimental projects to mission-critical, enterprise-wide deployments providing new insights, competitive advantage, and business innovation [23]. IDC estimates that the Big Data market grew from $3.2 billion in 2010 to $16.9 billion in 2015 with a compound annual growth rate of 39.4%, about seven times the one of the overall ICT market [7].

From the technological perspective, the MapReduce programming model is one of the most widely used solutions to support Big Data applications [27]. Its open source implementation, Apache Hadoop, is able to manage large datasets over either commodity clusters or high performance distributed topologies [45]. MapReduce has attracted the interest of both industry and academia, since analyzing large amounts of unstructured data is a high priority task for many companies and overtakes the scalability level that can be achieved by traditional data warehouse and business intelligence technologies [27]. Moreover, thanks to a wide set of performance enhancements (e.g., SSD support, caching, I/O barriers elimination) implemented within Hadoop 2.x, nowadays MapReduce can support both traditional batch and interactive data analysis applications [37].

---

* michele.ciavotta@polimi.it
** eugenio.gianniti@polimi.it
* * * danilo.ardagna@polimi.it

However, the adoption of Hadoop and other Big Data technologies is complex. The deployment and setup of an implementation is time-consuming, expensive, and resource-intensive. Companies need an *easy button* to accelerate the deployment of Big Data analytics [20]. For this reason, Cloud Computing is also becoming a mainstream solution to provide very large clusters on a pay-per-use basis. For example, Cloud storage provides an effective and cheap solution for storing Big Data. The pay-per-use approach and the almost infinite capacity of Cloud infrastructures can be used efficiently in supporting data intensive computations. Many Cloud providers already include in their offering MapReduce based platforms, among which Microsoft HDInsight [10] or Amazon Elastic MapReduce [2]. IDC estimates that, by 2020, nearly 40% of Big Data analyses will be supported by public Clouds [7], while Hadoop touched half of the world data last year [24].

In the very beginning, MapReduce jobs were meant to run on dedicated clusters to support batch analyses via a FIFO scheduler [35,36]. Nevertheless, MapReduce applications have evolved and nowadays large queries, submitted by different users, need to be performed on shared clusters, possibly with some guarantees on their execution time [48,50].

In this context one of the main challenges [28,41] is that the execution time of a MapReduce job is generally unknown in advance. In such systems, capacity allocation becomes one of the most important aspects. Determining the optimal number of nodes in a cluster shared among multiple users performing heterogeneous tasks is an important and difficult problem [41].

Our focus in this report is to provide a software tool able to support system administrators and operators in the capacity planning process of shared Hadoop 2.x Cloud clusters supporting both batch and interactive applications with deadline guarantees. Having such information available at design-time enables operators to make more informed decisions about the technology to use and to fully exploit the potential offered by the Cloud infrastructure.

We formulate the capacity planning problem by means of a mathematical model, with the aim of minimizing the cost of Cloud resources. The problem considers multiple virtual machine (VM) types as candidates to support the execution of Hadoop applications from multiple user classes. Cloud providers offer VMs of different capacity and cost. Given the complexity of virtualized systems and multiple bottleneck switches that occur during Big Data applications execution, very often the largest VM available is not the best choice from either the performance or performance/cost ratio perspective [49,19]. Through a search space exploration, our approach determines the optimal VM type and instances number, considering also specific Cloud provider pricing models (namely reserved and spot instances [1]). The underlying optimization problem is NP-hard and it is solved by a hill climbing heuristic able to determine the optimal configuration for a cluster governed by the YARN Capacity Scheduler.

Our work is one of the first contributions facing the problem of optimal sizing of Hadoop 2.x Cloud systems adopting the Capacity Scheduler [8]. Job execution times are estimated by relying on queueing network (QN) models, which are also a novel contribution of this report.

We validate the accuracy of our solutions on real systems by performing experiments based on the TPC-DS industry benchmark for business intelligence data warehouse applications [11]. Amazon EC2 and the CINECA Italian supercomputing center have been considered as target deployments. QN model simulation results and experiments performed on real systems have shown that the accuracy we can achieve is within 30% of the measurements in the worst case, with an average error around 12%.

This report is organized as follows. Section 2 presents in detail the problem addressed in the report. In Section 3 we focus on the formulation of the optimization problem and on the design-time exploration algorithm to solve it implemented by our D-SPACE4Cloud tool. In Section 4 we evaluate our approach by considering first the accuracy that can be achieved by our QN models and then the overall effectiveness of the optimization method. Finally, in Section 5 we compare our work with other proposals available in the literature and draw the conclusions in Section 6.

## 2    Problem Statement

In this section we aim at introducing some important details on the problem addressed in this work. We envision the following scenario, wherein a company needs to set up a cluster to carry out efficiently a set of interactive Big Data queries. A Hadoop 2.x cluster featuring the YARN Capacity Scheduler and running on a public Cloud IaaS is considered a fitting technological solution for the requirements of the company.

In particular, the cluster must support the parallel execution of Big Data applications in the form of Hadoop jobs or Hive/Pig queries. Different classes $\mathcal{C} = \{i \,|\, i = 1, \ldots, n\}$ gather applications that show a similar behavior. The cluster composition and size, in terms of type and number of VMs, must be decided in such a way that, for every application class $i$, $H_i$ jobs are guaranteed to execute concurrently and complete before a prearranged deadline $D_i$.

Moreover, YARN is configured in a way that all available cores can be dynamically assigned to either Map or Reduce tasks. Finally, in order to limit the risk of data corruption and according to the practices suggested by major Cloud vendors [2,10], the datasets reside on an external storage infrastructure [3,9] accessible at quasi-constant time.

As, in general, Infrastructure as a Service providers feature a limited, but possibly large, catalog of VM configurations $\mathcal{V} = \{j \,|\, j = 1, \ldots, m\}$ that differ in capacity (CPU speed, number of cores, available memory, etc.) and cost, making the right design-time decision poses a challenge that can lead to important savings throughout the cluster life-cycle. We denote with $\tau_i$ the VM type $j$ used to support jobs of class $i$ and with $\nu_i$ the number of VMs of such a kind allocated to class $i$. In this scenario, we consider a pricing model derived from *Amazon EC2* [1]. The provider offers: 1) *reserved* VMs, for which it adopts a one-time payment policy that grants access to a certain number of them for the contract duration; and 2) *spot* VMs, for which customers bid and compete for unused datacenter capacity, yielding very competitive hourly fees. In order to obtain the most cost-effective configuration, we rely on reserved VMs ($R_i$) for the bulk of computational needs and complement them with spot VMs ($s_i$). Let $\sigma_{\tau_i}$ be the unit cost for spot VMs of type $\tau_i$, whilst $\pi_{\tau_i}$ is the effective hourly cost for one reserved VM: it is the unit upfront payment normalized over the contract duration. Overall, the cluster hourly renting out costs can be calculated as follows:

$$\text{cost} = \sum_{i \in \mathcal{C}} (\sigma_{\tau_i} s_i + \pi_{\tau_i} R_i) \tag{1}$$

Let $\nu_i = R_i + s_i$: as the reliability of spot VMs depends on market fluctuations, to keep a high Quality of Service (QoS) the number of spot VMs is bounded not to be greater than a fraction $\eta_i$ of $\nu_i$ for each class $i$.

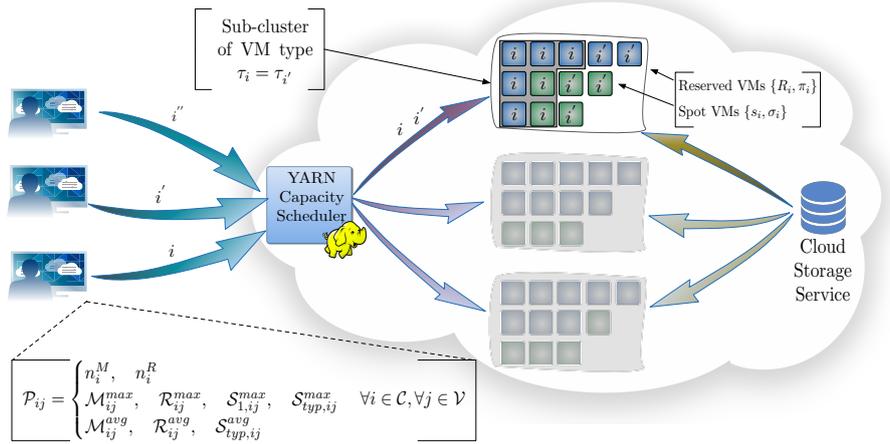

**Figure 1.** Reference system

Reducing the operating costs of the cluster by using efficiently the leased virtual resources is in the interest of the company. This translates into a Resource Provisioning problem where the renting out costs must be minimized subject to the fulfillment of QoS requirements, namely $H_i$ per-class concurrency level given certain deadlines $D_i$. In the following we assume that the system supports $H_i$ users for each class and that users work interactively with the system and run another job after a think time exponentially distributed with mean $Z_i$, i.e., the system is represented as a closed model [26]. In order to rigorously model and solve this problem, it is crucial to predict with fair confidence the execution times of each application class under different conditions: level of concurrency, cluster size, and composition. Following the approach presented in [41] it is possible to derive from the Hadoop logs a *job profile*, that is a concise behavior characterization for each class. Following the notation brought forth in [41,30], given a certain VM of type $j$, the job profile $\mathcal{P}_{ij}$ for application class $i$ aggregates the following information: 1) $n_i^M$ and $n_i^R$, respectively the total number of Map and Reduce tasks per job; 2) $\mathcal{M}_{ij}^{max}$, $\mathcal{R}_{ij}^{max}$, $\mathcal{S}_{1,ij}^{max}$, and $\mathcal{S}_{typ,ij}^{max}$, the maximum duration of a single Map, Reduce, and Shuffle task (notice that the first Shuffle wave of a given job is distinguished from all the subsequent ones); 3) $\mathcal{M}_{ij}^{avg}$, $\mathcal{R}_{ij}^{avg}$, and $\mathcal{S}_{typ,ij}^{avg}$, i.e., the average duration of Map, Reduce, and Shuffle tasks, respectively.

Given the amount and type of resources allocated, the concurrency level, and the job profile, the estimated execution time can generically be expressed as in (2):

$$T_i = \mathcal{T}\left(\mathcal{P}_{i,\tau_i}, \nu_i; H_i, Z_i\right), \quad \forall i \in \mathcal{C}. \tag{2}$$

What is worthwhile to note is that the previous formula represents a general relation describing either closed form results, as those presented in [30], or the average execution times derived via simulation, the approach adopted in this report. Since the execution of jobs on a suboptimal VM type might give rise to performance disruptions, it is critical to avoid assigning tasks belonging to class $i$ to the wrong VM type $j \neq \tau_i$. Indeed, YARN allows for specifying Node Labels and partitioning nodes in the cluster according to these labels, then it is possible to enforce this separation. Our configuration statically splits different VM types with this mechanism and adopts within each partition either a further static separation in classes or a work conserving scheduling mode, where idle resources can be assigned to jobs requiring the same VM type. The assumption on the scheduling policy governing the exploitation of idle resources is not critical: it only affects the interpretation of results, where the former case leads to sharp predictions, while in the latter the outcomes of the optimization algorithm are upper bounds, with possible performance improvements due to a better cluster utilization. Equations (2) can be used to formulate the deadline constraints as:

$$T_i \leq D_i, \quad \forall i \in \mathcal{C}. \tag{3}$$

In light of the above, we can say that the ultimate goal of the proposed approach is to determine the optimal VM type selection $\tau_i$ and number and pricing models of VMs $\nu_i = R_i + s_i$ for each class $i$ such that the sum of costs is minimized, while the deadlines and concurrency levels are met.

The reader is referred to Figure 1 for a graphical overview of the main elements of the considered resource provisioning problem. Furthermore, in Table 1 a complete list of the parameters used in the models presented in the next sections is reported, whilst Table 2 summarizes the decision variables.

**Table 1.** Model parameters

| Parameter | Definition |
|---|---|
| $\mathcal{C}$ | Set of application classes |
| $\mathcal{V}$ | Set of VM types |
| $H_i$ | Number of concurrent users for class $i$ |
| $Z_i$ | Class $i$ think time [ms] |
| $D_i$ | Deadline associated to applications of class $i$ [ms] |
| $\eta_i$ | Maximum percentage of spot VMs allowed to class $i$ |
| $\sigma_j$ | Unit hourly cost for spot VMs of type $j$ [€/h] |
| $\pi_j$ | Effective hourly price for reserved VMs of type $j$ [€/h] |
| $\mathcal{P}_{ij}$ | Job profile of class $i$ with respect to VM type $j$ |

**Table 2.** Decision variables

| Variable | Definition |
|---|---|
| $\nu_i$ | Number of VMs assigned for the execution of applications from class $i$ |
| $R_i$ | Number of reserved VMs booked for the execution of applications from class $i$ |
| $s_i$ | Number of spot VMs assigned for the execution of applications from class $i$ |
| $x_{ij}$ | Binary variable equal to 1 if class $i$ is hosted on VM type $j$ |

## 3 Problem Formulation and Solution

In the following we present the optimization model and techniques exploited by the D-SPACE4Cloud tool in order to determine the optimal VM mix given the profiles characterizing the applications under study and the possible Cloud providers to host the virtual cluster, as well as the QN models used to obtain an accurate assessment of the execution times given the resource configuration. Further, we describe the heuristic algorithm adopted to efficiently tackle the resource provisioning problem by exploiting the presented models.

### 3.1 Optimization and Performance Models

Basic building blocks for this tool are the models of the system under study. First of all, we need a quick, although rough, method to estimate completion times and operational costs: to this end, we exploit a mathematical programming formulation. In this way, it is possible to swiftly explore several possible configurations and point out the most cost-effective among the feasible ones. Afterwards, the required resource configuration can be fine-tuned using more accurate, even if more time consuming and computationally demanding, QN simulations, reaching a precise prediction of the expected response time.

According to the previous considerations, the first step in the optimization procedure consists in determining the most cost-effective resource type, based on their price and the expected performance. This will be done by exploiting a set of logical variables $x_{ij}$: we will enforce that only $x_{i,\tau_i} = 1$, thus determining the optimal VM type $\tau_i$ for application class $i$. We address this issue proposing the following mathematical programming formulation:

$$\min_{\mathbf{x},\boldsymbol{\nu},\mathbf{s},\mathbf{R}} \sum_{i \in \mathcal{C}} (\sigma_{\tau_i} s_i + \pi_{\tau_i} R_i) \tag{P1a}$$

subject to:

$$\sum_{j \in \mathcal{V}} x_{ij} = 1, \quad \forall i \in \mathcal{C} \tag{P1b}$$

$$\mathcal{P}_{i,\tau_i} = \sum_{j \in \mathcal{V}} \mathcal{P}_{ij} x_{ij}, \quad \forall i \in \mathcal{C} \tag{P1c}$$

$$\sigma_{\tau_i} = \sum_{j \in \mathcal{V}} \sigma_j x_{ij}, \quad \forall i \in \mathcal{C} \tag{P1d}$$

$$\pi_{\tau_i} = \sum_{j \in \mathcal{V}} \pi_j x_{ij}, \quad \forall i \in \mathcal{C} \tag{P1e}$$

$$x_{ij} \in \{0,1\}, \quad \forall i \in \mathcal{C}, \forall j \in \mathcal{V} \tag{P1f}$$

$$(\boldsymbol{\nu}, \mathbf{s}, \mathbf{R}) \in \arg\min \sum_{i \in \mathcal{C}} (\sigma_{\tau_i} s_i + \pi_{\tau_i} R_i) \tag{P1g}$$

subject to:

$$s_i \leq \frac{\eta_i}{1-\eta_i} R_i, \quad \forall i \in \mathcal{C} \tag{P1h}$$

$$\nu_i = R_i + s_i, \quad \forall i \in \mathcal{C} \tag{P1i}$$

$$\mathcal{T}(\mathcal{P}_{i,\tau_i}, \nu_i; H_i, Z_i) \leq D_i, \quad \forall i \in \mathcal{C} \tag{P1j}$$

$$\nu_i \in \mathbb{N}, \quad \forall i \in \mathcal{C} \tag{P1k}$$

$$R_i \in \mathbb{N}, \quad \forall i \in \mathcal{C} \tag{P1l}$$

$$s_i \in \mathbb{N}, \quad \forall i \in \mathcal{C} \tag{P1m}$$

Problem (P1) is a bilevel resource allocation problem where the outer objective function (P1a) considers running costs. The first set of constraints, (P1b), associates each class $i$ with only one VM type $j$, hence the following constraints, ranging from (P1c) to (P1e), pick the values for the inner problem parameters.

The inner objective function (P1g) has the same expression as (P1a), but in this case the prices $\sigma_{\tau_i}$ and $\pi_{\tau_i}$ are fixed, as they have been chosen at the upper level. The following constraints, (P1h), enforce that spot instances do not exceed a fraction $\eta_i$ of the total assigned VMs and constraints (P1i) add all the VMs available for class $i$, irrespective of the pricing model. Further, constraints (P1j) mandate to respect the deadlines $D_i$, as stated in the Service Level Agreement contracts. In the end, all the remaining decision variables are taken from the natural numbers set, according to their interpretation.

The presented formulation of Problem (P1) is particularly difficult to tackle, as it is a bilevel mixed integer nonlinear programming (MINLP) problem, possibly nonconvex, depending on $\mathcal{T}$. According to the literature about complexity theory [18], integer programming problems belong to the NP-hard class, hence the same applies to (P1). However, since there is no constraint linking variables belonging to different application classes, we can split this general formulation into several smaller and independent problems, one per class $i$.

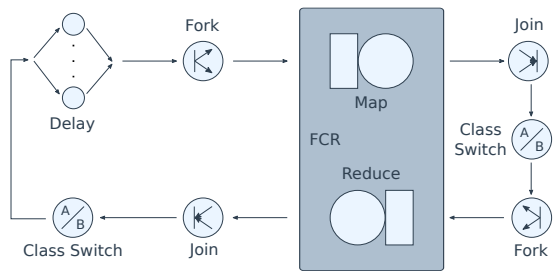

**Figure 2.** Queueing network model

We still have the problem of computing the average completion time $T$. In this work, we exploit a QN model for MapReduce jobs, assuming that the YARN Capacity Scheduler is used. In the following, we do not make any assumptions on the configuration of the Capacity Scheduler, according to the considerations presented in Section 2. The model can be applied to both statically partitioned and work conserving mode clusters, but care should be taken in the interpretation of results. In the former case, our model provides the mean completion time for every job class. On the other hand, if the scheduler is configured in work conserving mode, then the completion time we obtain is an upper bound, due to possible performance gains when resources are exploited by other classes instead of lying idle. Exploiting the node labeling mechanism, we enforce that jobs run on their optimal VM type $\tau_i$ in any case, so as to avoid cases where some tasks running on improper nodes become bottlenecks and disrupt both the lender and borrower class performance.

The performance model is depicted in Figure 2. It is a closed QN model where the number of concurrent users is given by $H_i$ and they start off in the delay center characterized by $Z_i$. When a user submits their job, this is forked into as many Map task requests as stated in the job profile $\mathcal{P}_{i,\tau_i}$, which then enter the finite capacity region (FCR) [15]. FCRs model situations where several service centers access resources belonging to a single limited pool, competing to use them. Hence, the FCR enforces an upper bound on the total number of requests served at the same time within itself, allowing tasks in based on a FIFO queue and supporting prioritization of different classes. The FCR includes two multi-service queues that model the Map and Reduce execution stages. FCR and multi-service queues capacities are equal to the total number of cores available to class $i$. In this way, we can model the dynamic assignment of YARN containers to Map and Reduce tasks whenever they are ready.

Map tasks are executed by the first multi-service queue and synchronize after completion by joining back to a single job request. The Reduce phase is modeled analogously, while between the phases there are class switches to enforce that Reduce tasks waiting for resources obtain them with priority. Indeed, the YARN Capacity Scheduler implements a FIFO scheduling within the same queue and containers are allocated to the next job only when all Reduce tasks have obtained enough resources. Reduce tasks have a higher priority than Map tasks to enforce this behavior. Note that the Map join is external to the FCR in order to model that when Map tasks complete they release container cores, which can be assigned to tasks ready in the FCR FIFO queue. Moreover, the Reduce fork is also external to the FCR to model correctly applications characterized by a number of Reducers larger than the total number of cores. Note that the model in Figure 2 is rather general and can be easily extended to consider also Tez or Spark applications, where a Tez directed acyclic graph (DAG) node or Spark stage is associated to a corresponding multi-server queue.

### 3.2 Solution Techniques

The aim of this section is to provide a brief description of the optimization approach embedded in D-SPACE4Cloud. The tool implements an optimization mechanism that efficiently explores the space of possible configurations.

Figure 3 depicts the main elements of the D-SPACE4Cloud architecture that come into play in the optimization scenario. The tool takes as input a description of the considered problem, consisting in a set of applications, a set of suitable VMs for each application along with the respective job profiles for each machine, and QoS constraints expressed in terms of deadlines for each considered application. Specifically, all these parameters are collected in a JSON file provided as input to the tool. The *Initial Solution Builder* generates a starting solution for the problem using a MINLP formulation where the time expression $\mathcal{T}$ appearing in constraint (P1j) is a convex function: then the inner level of (P1) is a convex nonlinear problem and we exploit the Karush-Kuhn-Tucker conditions to speed up the solution process. For more details the reader is referred to [29]. It must be highlighted, at this point, that the quality of the returned solution can still be improved: this because the MINLP relies on an approximate representation of the Application-Cluster liaison. For this reason the QN model presented in the previous section is exploited to get a more accurate execution time assessment. The increased accuracy leaves room for further cost reduction; however, since QNs are time consuming tools, the space of possible cluster configurations has to be explored in the most efficient way, avoiding to evaluate unpromising configurations.

In the light of such considerations, a heuristic approach has been adopted and a component called *Parallel Local Search Optimizer* has been devised. Internally, it implements a parallel hill climbing (HC) technique to optimize the number of replicas of the assigned resource for each application; the goal is to find the minimum number of resources to fulfill the QoS requirements. This procedure is applied independently, and in parallel, on all application classes and terminates when a further reduction in the number of replicas would lead to an infeasible solution. As soon as all the classes reach convergence, it is possible to retrieve from the D-SPACE4Cloud tool a JSON file listing the results of the overall optimization procedure. In particular, HC is a local-search-based procedure that operates on the current solution performing a change (more often referred to as *move*) in the structure of the solution in such a way that the newly generated solution could possibly show an improved objective value. If the move is successful it is applied again on the new solution and the process is repeated until no further improvement is possible. The HC algorithm stops when a local optimum is found; however, if the objective to optimize is convex, HC is able to find the global optimum. This is the case of the considered cost function (1), which is linear in the number of VMs in the cluster, since VM prices are fixed at the first level. Hence, every feasible instance of the inner problem can be heuristically solved to optimality through HC.

Algorithm 1 is reported here for clarity purposes. The initial solution $S$, obtained from the MINLP solution, is evaluated using the QN model and each one of its parts is optimized separately and in parallel (line 2). If the partial solution $S_i$ is infeasible the size of its personal cluster is increased by one unit (line 5) until it reaches feasibility. Otherwise, the procedure attempts to decrease the cost function by reducing the cluster size (line 10). Finally, it is worth pointing out that every time the total number of machines in a cluster is incremented or decremented the best mix of pricing models (i.e., $R_i$, $s_i$) is computed so as to minimize the renting out costs of that configuration.

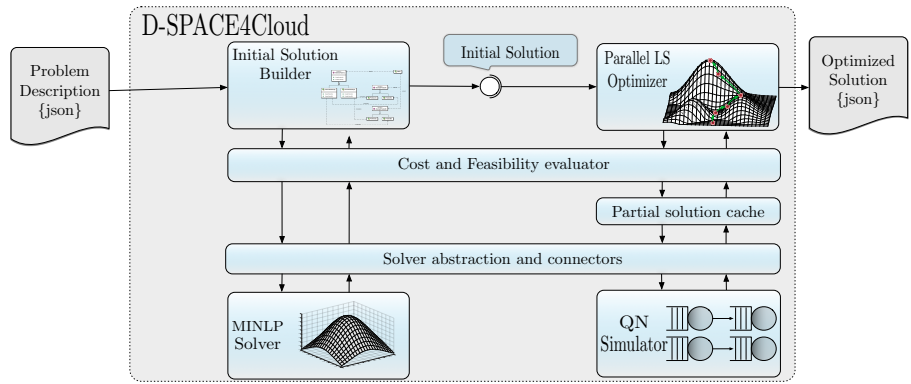

**Figure 3.** D-SPACE4Cloud architecture

**Algorithm 1** Hill climbing algorithm
---
**Require:** $S = \{S_i \,|\, i \in \mathcal{C}\}$
1: ***Evaluate*** $(S)$
2: **for all** $i \in \mathcal{C}$ **do** ← *Parallel for*
3:     **if** $S_i$ is infeasible **then**
4:         **while** $S_i$ is infeasible **do** ⎫
5:            ***IncrementCluster*** $(S_i)$ ⎬ *Pursuit of feasibility*
6:            ***Evaluate*** $(S_i)$ ⎭
7:         **end while**
8:     **else**
9:         **while** $S_i$ is feasible **do** ⎫
10:           ***DecrementCluster*** $(S_i)$ ⎬ *Cost optimization*
11:           ***Evaluate*** $(S_i)$ ⎭
12:         **end while**
13:         ***IncrementCluster*** $(S_i)$
14:     **end if**
15: **end for**
16: **return** $S$

## 4 Experimental Analysis

In this section we show the results of several experiments performed to validate the proposed approach. All these experiments have been performed on two Ubuntu 14.04 VMs hosted on an Intel Xeon E5530 2.40 GHz equipped server. The first VM ran the D-SPACE4Cloud web service and KNITRO 10.0 [6], a solver for the mathematical programming language AMPL [4], which was used to solve the optimization problem presented in Section 3.1 (see [29] for further details), determining an initial solution to our HC algorithm. The second one, instead, ran JMT 0.9.3 [15], a collection of free software performance evaluation programs including a QN simulator.

### 4.1 Experimental Setup and Design of Experiments

In order to obtain job profiles, we chose a set of SQL queries, shown in Figure 4, from the industry standard benchmark TPC-DS [11]. We then generated synthetic data compliant with the specifications and executed the queries on Apache Hive [5]. Notice that we generated data at several scale factors ranging from 250 GB to 1 TB. Since profiles collect statistical information about jobs, we repeated the profiling runs at least twenty times per query. Properly parsing the logs allows to extract all the parameters composing every query profile, for example average and maximum task execution times, number of tasks, etc. Profiling has been performed on Amazon EC2, by considering m4.xlarge instances, and on PICO[1], the Big Data cluster offered by CINECA, the Italian supercomputing center. The cluster rented on EC2 was composed of 30 computational

---
[1] http://www.hpc.cineca.it/hardware/pico

```
select avg(ws_quantity),
       avg(ws_ext_sales_price),
       avg(ws_ext_wholesale_cost),
       sum(ws_ext_wholesale_cost)
from web_sales
where (web_sales.ws_sales_price between 100.00 and 150.00) or (web_sales.
    ws_net_profit
between 100 and 200)
group by ws_web_page_sk
limit 100;
```

(a) Q1

```
select inv_item_sk,inv_warehouse_sk
from inventory where inv_quantity_on_hand > 10
group by inv_item_sk,inv_warehouse_sk
having sum(inv_quantity_on_hand)>20
limit 100;
```

(b) Q2

```
select avg(ss_quantity), avg(ss_net_profit)
from store_sales
where ss_quantity > 10 and ss_net_profit > 0
group by ss_store_sk
having avg(ss_quantity) > 20
limit 100;
```

(c) Q3

```
select cs_item_sk, avg(cs_quantity) as aq
from catalog_sales
where cs_quantity > 2
group by cs_item_sk;
```

(d) Q4

```
select inv_warehouse_sk, sum(inv_quantity_on_hand)
from inventory
group by inv_warehouse_sk
having sum(inv_quantity_on_hand) > 5
limit 100;
```

(e) Q5

**Figure 4.** Queries

nodes, for a total of 120 vCPUs hosting 240 containers, whilst on PICO we used up to 120 cores configured to host one container per core. In the first case every container had 2 GB RAM and in the second 6 GB. Along with profiles, we also collected lists of task execution times to feed into the replayer in JMT service centers. In the end, we recorded the different VM types characteristics.

### 4.2 Queueing Network Validation

To start off with, we show results for the validation of the QN model shown in Figure 2. We feed the model with parameters evaluated on the real systems we took into account and compare the measured performance metrics with the ones obtained via simulation. Specifically, we consider as a quality index the accuracy on the prediction of response times, defined as follows:

$$\vartheta = \frac{\tau - T}{T} \qquad (4)$$

where $\tau$ is the simulated response time, whilst $T$ is the average measured one. Such a definition allows not only to quantify the relative error on response times, but also to identify cases where the predicted time is smaller than the actual one, thus leading to possible deadline misses. Indeed, if $\vartheta < 0$ then the prediction is not conservative.

Among these experiments, we considered both single user scenarios, where one query has been run repeatedly on dedicated cluster, interleaving a 10 s average think time between completions and subsequent submissions, and multiple user scenarios, with several users concurrently interacting with the cluster in a similar way.

**Table 3.** Queueing network model validation

| Query | Users | Cores | Dataset [GB] | $n^M$ | $n^R$ | $T$ [ms] | $\tau$ [ms] | $\vartheta$ [%] |
|---|---|---|---|---|---|---|---|---|
| Q1 | 1 | 240 | 250 | 500 | 1 | 55,410 | 50,240.96 | −9.33 |
| Q1 | 5 | 40 | 250 | 144 | 151 | 637,888 | 808,330.61 | 26.72 |
| Q2 | 1 | 240 | 250 | 65 | 5 | 36,881 | 28,022.81 | −24.02 |
| Q2 | 3 | 20 | 250 | 4 | 4 | 95,403 | 92,581.74 | −2.96 |
| Q3 | 1 | 240 | 250 | 750 | 1 | 76,806 | 77,719.30 | 1.19 |
| Q4 | 1 | 240 | 250 | 524 | 384 | 92,197 | 76,956.52 | −16.53 |
| Q1 | 1 | 60 | 500 | 287 | 300 | 378,127 | 411,940.93 | 8.94 |
| Q3 | 1 | 100 | 500 | 757 | 793 | 401,827 | 524,759.36 | 30.59 |
| Q3 | 1 | 120 | 750 | 1,148 | 1,009 | 661,214 | 759,230.77 | 14.82 |
| Q4 | 1 | 60 | 750 | 868 | 910 | 808,490 | 844,700.85 | 4.48 |
| Q3 | 1 | 80 | 1,000 | 1,560 | 1,009 | 1,019,973 | 1,053,829.78 | −1.00 |
| Q5 | 1 | 80 | 1,000 | 64 | 68 | 39,206 | 36,598.32 | −6.65 |

Table 3 shows the results of the QN model validation. In the worst case, the relative error can reach up to 30.59%, which is perfectly in line with the expected accuracy on response time prediction [26]. Instead, overall the average relative error settles at 12.27%.

### 4.3 Scenario-based Experiments

The optimization approach described in Section 3 needs to be validated, ensuring that it is capable of catching realistic behaviors we expect of the system under analysis. We test this property with a set of assessment runs where we fix all the problem parameters but one and verify that the solutions follow an intuitive evolution.

The main axes governing performance in Hadoop clusters hosted on public Clouds are the level of concurrency and the deadlines. In the first case, increasing $H_i$ and fixing all the remaining parameters, we expect a need for more VMs to support the rising workload, thus leading to an increase of renting out costs. On the other hand, if at fixed parameters we tighten the deadlines $D_i$, again we should observe increased costs: the system will require a higher parallelism to shrink response times, hence more computational nodes to support it.

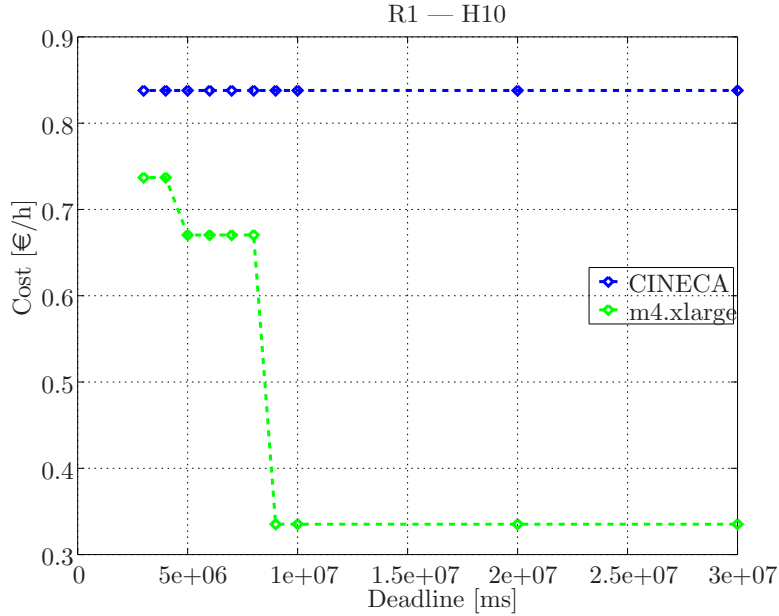

**Figure 5.** Query R1, 10 concurrent users

For the sake of clarity, we performed single-class experiments: considering only one class per experiment allows for an easier interpretation of the results. Figures 5, 6 and 7 report the solutions obtained with the 250 GB dataset profiles.

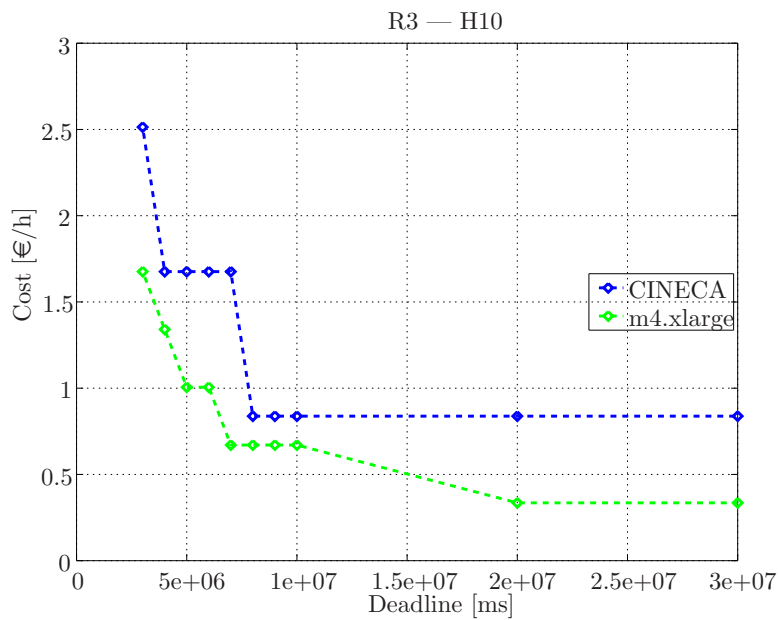

**Figure 6.** Query R3, 10 concurrent users

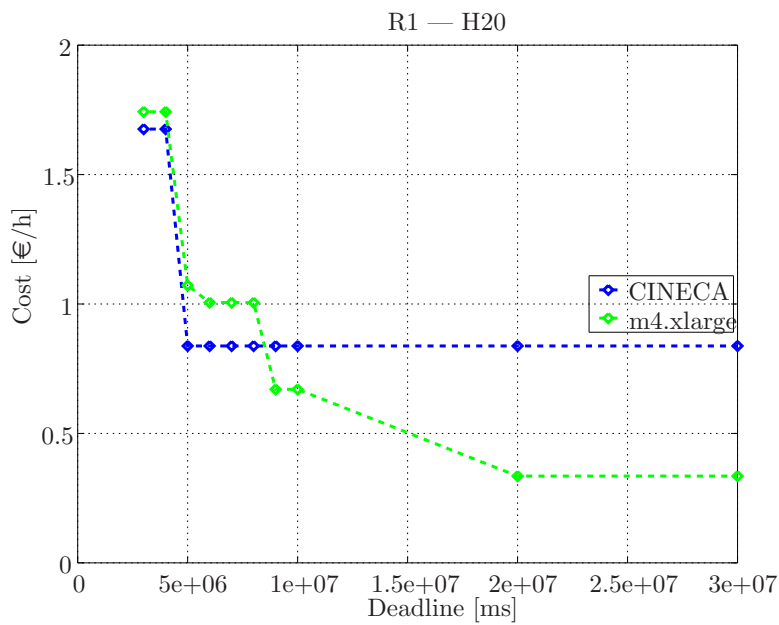

**Figure 7.** Query R1, 20 concurrent users

The average running time for these experiments is about two hours. All the mentioned figures show the cost in €/h plotted against decreasing deadlines in ms for both the real VM types considered: CINECA is the 20-core node available on PICO, whilst m4.xlarge is the 4-core instance rented on Amazon AWS. In Figures 5 and 6 the expected cost increase due to tightening deadlines is apparent for both query R1 and R3, considering 10 concurrent users. Further, in both cases it is cheaper to provision a Cloud cluster consisting of the smaller Amazon-offered instances, independently of the deadlines. It is then interesting to observe that R1 shows a different behavior if the required concurrency level increases. Figure 7 shows that, as the deadlines become tighter and tighter, it is possible to identify a region where executing the workload on larger VMs becomes more economic.

## 5 Related Work

Capacity planning and architecture design space exploration are important problems analyzed in the literature [12,16]. High level models and tools to support software architects (see, e.g., Palladio Component Model and its Palladio Bench and PerOptirex design environment [14,25], or stochastic process algebra and the PEPA Eclipse plugin [40,33]) have been proposed for identifying the best configuration given a set of QoS requirements for enterprise web-based systems, but unfortunately they do not support Cloud-specific abstractions or do not directly address the problem of deriving an optimized Cloud and Big Data cluster configuration. On the other side, capacity management and cluster sizing for Big Data applications has received also a widespread interest by both academia and industry.

The starting point is the consideration that Hadoop often requires an intense tuning phase in order to exhibit its full potential. For this reason, *Starfish*, a self-tuning system for analytics on Hadoop, has been proposed [22]. In particular, Starfish collects some key runtime information about the applications execution with the aim of generating meaningful application profiles; such profiles are in turn the basic elements to be exploited for Hadoop automatic configuration processes. Furthermore, also the cluster sizing problem has been tackled and successfully solved exploiting the same tool [21].

The resource provisioning problem, instead, has been faced by Tian and Chen [39]. The goal is the minimization of the execution cost for a single application. They present a cost model that depends on the dataset size and on some characteristics of the considered application. A regression-based analysis technique has been used to profile the application and to estimate model parameters. The problem of job profiling and execution time estimation represents a common issue in the Big Data literature. Verma et al. [42] proposed a framework for the profiling and duration prediction of applications running on heterogeneous resources. An approach to this problem based on closed queueing networks is presented in [13]. This work is noteworthy as it explicitly considers contention and parallelism on compute nodes to evaluate the execution time of a MapReduce application. However, the weak spot of this approach is that it contemplates the Map phase alone. Vianna et al. [43] worked on a similar solution; however the validation phase has been carried out considering a cluster dedicated to the execution of a single application at a time.

Both Map and Reduce phases are considered in [38]. In this work the Map phase is modeled as an M/G/1 queue, whereas for the Reduce phase a multi-server queue have been used. Note, however, that this approach has been especially tailored for Reduce-intensive applications. As far as queue models are concerned, the case of Lin et al. [28] deserves to be cited; the authors propose to model the execution of the application by means of a tandem queue with overlapping phases. In the same work, efficient runtime scheduling solutions for the joint optimization of the Map and copy/shuffle phases are provided. The authors demonstrated the effectiveness of their approach comparing it with the offline-generated optimal schedule.

Castiglione et al. [17] introduce a novel modeling approach based on mean field analysis and provide fast approximate methods to predict the performance of Big Data systems.

Deadlines for MapReduce jobs are considered in [34]. The authors recognize the inadequacy of Hadoop schedulers released at the date to properly handle completion time requirements. The work proposes to adapt to the problem some classical multiprocessor scheduling policies; in particular, two versions of the Earliest Deadline First heuristic are presented and proved to outperform off-the-shelf schedulers. A similar approach is presented in [46], where the authors present a solution to manage clusters shared among Hadoop application and more traditional Web systems.

The problem of progress estimation of multiple parallel queries is addressed in [32]. To this aim, the authors present Parallax, a tool able to predict the completion time of MapReduce jobs. The tool has been implemented over Pig, while the PigMix benchmark has been used for the evaluation. ParaTimer [31], an extension of Parallax, features support to multiple parallel queries expressed as DAGs.

Zhang et al. [47] investigate the performance of MapReduce applications on homogeneous and heterogeneous Hadoop clusters in the Cloud. They consider a problem similar to ours and provide a simulation-based framework for minimizing cluster infrastructural costs. Nonetheless, a single class of workload is optimized. The work in [44] faced the problem of optimization of mixed interactive and batch applications in system-on-a-chip solutions with heterogeneous cores.

In [41] the ARIA framework is presented. This work is the closest to our contribution and focuses on clusters dedicated to single user classes handled by the FIFO scheduler. The framework addresses the problem of calculating the most suitable number of resources to allocate to Map and Reduce tasks in order to meet a user-defined due date for a certain application; the aim is to avoid as much as possible costs due to resource over-provisioning. We borrow from this work the compact job profile definition, used there to calculate a lower bound, an upper bound, and an estimation of an application execution time. Finally, a performance model is presented, and eventually improved in [50], and then validated through a simulation study and an experimental campaign on a 66-node Hadoop cluster. The same authors, in a more recent work [48], provided a solution for optimizing the execution of a workload specified as a set of DAGs under the constraints of a global deadline or budget. Heterogeneous clusters with possible faulty nodes are considered as well.

## 6 Conclusions

In this report we have proposed a novel approach to provisioning Cloud clusters to support data intensive applications, specifically SQL queries executed with Hive over Hadoop YARN managed clusters. Building upon a performance model available in the technical literature, we have developed a MINLP formulation and a QN model to assess the overall performance of Big Data systems, knowing the involved applications job profiles and the resource configuration of the underlying cluster. In order to achieve a favorable trade-off between prediction accuracy and running times, we have adopted a heuristic approach that exploits the fast solvers available for mathematical programming problems for the initial exploration of the solution space and then relies on the precise, if slower, QN techniques, thus getting the best of both worlds.

Moreover, our experimental validation shows how our tool is a valuable contribution towards supporting different application classes over heterogeneous VM types, since we have highlighted situations where sticking to small instances and scaling out proves to be less economic than switching to better equipped VMs that allow for a smaller number of replicas: the decreased replication factor compensates the increased unit price in a not obvious way.

Moving from the presented results, an interesting research direction for our future work lies in the characterization of complex workflows expressed as DAGs, e.g., Tez or Spark jobs. Another relevant aspect to investigate is the usage of more sophisticated techniques for the heuristic exploration of the solution space, in order to attain further speedup and, possibly, extend our method to the runtime cluster management scenario.